\documentstyle[12pt]{article}
\begin{document}
\title{TOPOLOGICAL ASPECTS OF THE NON-ADIABATIC BERRY PHASE}
\author{Ali MOSTAFAZADEH \thanks{Center for Relativity} \and Arno BOHM
\thanks{Center for Particle Theory} \\
Department of Physics \\
The University of Texas at Austin \\ Austin , Texas 78712}
\maketitle
\baselineskip=18pt
\begin{abstract}
The topology of the non-adiabatic parameter space bundle is discussed
for evolution of exact cyclic state vectors in
Berry's original example of split angular momentum eigenstates.
It turns out that the change in topology occurs
at a  critical frequency. The first Chern number that classifies these
bundles is proportional to angular momentum. The non-adiabatic principal
bundle over the parameter space is not well-defined at the critical
frequency.
\end{abstract}
\newpage
\section{INTRODUCTION}

In a preceding paper,
\cite{ali-bohm} the relation between the parameter space and
the projective space approach to the problem of geometric (Berry) phase
is discussed.
The key idea is to use the classification theorem for principal bundles
or alternatively for vector bundles. This method suggests a way of
constructing fibre bundles over parameter space , even for the non-adiabatic
evolution of state vectors (at least for a particular class of
quantum systems).

Let $ M $ denote the smooth , compact \footnote{ In most physical examples
$M$ is compact. However, compactness is not a necessary condition in a
large part of our analysis. As far as the mathematical results are concerned,
$M$ must at least be paracompact.} manifold of parameters $ x $
, and $ H = H(x) $ be a Hamiltonian which depends smoothly on $ x \in M.$
As described in \cite{ali-bohm} , the cyclic state vectors are not energy
eigenvectors. The adiabaticity assumption, therefore does not describe the
actual situation, it provides just an approximation. One has to distinguish
two regions: the adiabatically related region for which the adiabatcity
assumption yields a limiting case, and the non-adiabatic region.
In the non-adiabatic region, one can not use the adiabatic theorem and
the cyclic state vectors can not be approximated by the eigenvectors of
the Hamiltonian.

The relation between Berry-Simon (B-S) \cite{simon} parameter space
bundle interpretation of the geometric phase, and the  Aharonov-Anandan
(A-A) \cite{aharonov-anandan} approach of using the projective space
bundle, is the following: The former principal bundle:
\newcommand{\be}{\begin{equation}}
\newcommand{\ee}{\end{equation}}
\newcommand{\okk}{\longrightarrow}
\newcommand{\lokk}{\longleftarrow}
\newcommand{\sokk}{\rightarrow}
\be
\lambda _{\cal N} : G \okk ... \okk M
\ee
is obtained as a pullback bundle from the latter:
\be
\eta _{\cal N} : G \okk V_{\cal N} \okk Gr_{\cal N}
\ee
In (1) and (2) , $ G = U({\cal N}) $ or $ O({\cal N}) $ depending on whether
the Hilbert space is real or complex , ${\cal N} < \infty $ is the dimension
of the degeneracy subspace where the cyclic state vector belongs ,
$ V_{\cal N} $ ,and $ Gr_{\cal N} $ are infinite dimensional (real or
complex) Stiefel and Grassmann manifolds , respectively \cite{ali-bohm}.
In other words, there is a continuous map $ f:M \sokk Gr_{\cal N} $ such
that the following diagram commutes:
\be
\begin{array}{ccc}
\lambda _{\cal N} & \stackrel{f^{\star}}{\lokk} & \eta _{\cal N} \\
\downarrow & \bigcirc & \downarrow \\
M & \stackrel{f}{\okk} & Gr_{\cal N}
\end{array}
\ee
and $f^{\star}$ is a bundle isomorphism , i.e.\ $ \lambda _{\cal N} \cong
f^{\star}(\eta _{\cal N})$.

For the adiabatic case $f$ is given by the Hamiltonian, namely:
\be
\forall x \in M \hspace*{1in} f(x) \equiv \mid \psi _{x}><\psi _{x} \mid
\ee
where $ \mid \psi _{x}><\psi _{x} \mid $ is the eigenstate of $H(x)$
which evolves in time as $x$ traverses a closed loop:
\be
C:[0,T] \ni t \okk x(t) \in M \hspace{0.2in},\hspace{0.2in} x(0) = x(T).
\ee

To repeat the same construction for the non-adiabatic case , one needs
to consider a class of quantum systems , for which:
\begin{enumerate}
\item The cyclic state vectors exist and are eigenvectors of an operator
 \mbox{$ \tilde{H}= \tilde{H}(x) $}.
\item $ \tilde{H}(x)=H(F(x))$ , for some continuous (smooth) function
$ F:M \sokk M $.
\end{enumerate}
For this class of quantum systems , one can still use the classification
theorem. This is realized by replacing $f$ by a map $ \tilde{f} $
defined by :
\[ \tilde{f} \equiv foF :M \okk Gr_{\cal N}. \]
Then the exact (non-adiabatic) bundle $\tilde{\lambda}_{\cal N}$ over
the parameter space is obtained as the pullback bundle
$ \tilde{\lambda}_{\cal N} = \tilde{f}^{\star}(\eta _{\cal N})$.

$f$ and $\tilde{f}$ pullback the canonical connection of the universal
bundles $\eta _{\cal N}$ onto $\lambda _{\cal N}$ and
$\tilde{\lambda}_{\cal N}$ , and yield the adiabatic and non-adiabatic
Berry connections , respectively. The geometric (Berry) phase is
identified with the holonomy of these connections , in each case.

The topologies of $\lambda _{\cal N}$ and $ \tilde{\lambda}_{\cal N}$
depend only on the homotopy classes of $\left[ f \right]$ and
$\left[ \tilde{f} \right] $ in $[M,Gr_{\cal N}]$. Hence , if $F$ is
a diffeomorphism \footnote{Note that in the adiabatic limit $\tilde{f}$
approaches to $f$ , hence, if $F$ is a diffeomorphism it will be
necessarily homotopic to identity.
In other words, $F$ as an element of $Diff(M)$ belongs to the
connected component to the identity.}(homotopic to identity will
suffice) $[f]=[\tilde{f}]$ , and $\lambda _{\cal N} \cong
\tilde{\lambda}_{\cal N}$.
\def \inbar{\vrule height1.5ex width.4pt depth0pt}
\def \IC{\relax\hbox{\kern.25em$\inbar\kern-.3em{\rm C}$}}
\def \IR{\relax{\rm I\kern-.18em R}}
\newcommand {\iz}{\ Z \hspace{-.08in}Z}

For the non-degenerate (abelian) case, ${\cal N}=1$ ($G=U(1)$) and
\[ Gr_{1}=\IC P(\infty )=K(2,\iz ) \] is an Eilenberg-McLane space.
This allows one to have the following one-to-one correspondence:
\[  [M,\IC P(\infty )] \cong H^{2}(M,\iz ). \]
Hence , it is the first Chern class $c_{1} \in H^{2}(M,\iz )$ that
determines the topology of $U(1)$ bundles ,
$\lambda \equiv \lambda _{1}$ and $ \tilde{\lambda} \equiv
\tilde{\lambda}_{1}$.

In the present paper , we study a particular example for which the
assumptions 1. and 2. are valid \footnote{Except for one instance}.
Thus , we explore the topology of the bundles $\lambda $ and
$\tilde{\lambda}$ , and discuss the rather interesting implications
of the topological information.

\section{NON-ADIABATIC EVOLUTION IN BERRY'S EXAMPLE}

We consider a magnetic dipole $\vec{\mu}$ in a magnetic field $B(x(t))$.
The Hamiltonian of this system is given by \cite{berry} :
\be
H(x) = - \vec{\mu}. \vec{B}(x) = b \vec{x}. \vec{J} \hspace*{.2in},
\hspace*{.2in} b=\frac{Bge}{2mc}.
\ee
The parameter space is $S^{2}$ and the closed loops considered are
the following circular loops:
\be
C:[0,T] \ni t \okk \vec{x}(t)=(\theta =const.,\varphi =
\omega t) \in S^{2}
\ee
where $\vec{x} \in S^2 \subset \IR ^{3}$ and
$(\theta , \varphi )$ are spherical coordinates with
respect to a standard Cartesian coordinate basis $(\hat{e_{1}} ,
\hat{e_{2}} ,\hat{e_{3}})$ or $(\hat{i} ,\hat{j} ,\hat{k} )$ of
$\IR ^{3}$.

The Schr\"{o}dinger equation is exactly solvable in this case
\cite{rabi,bohm}. The Hamiltonian $H$ can be written in the following
form:
\be
H=U^{\dagger}H_{0}U
\ee
where \[ U \equiv \exp (\frac{it\omega}{\hbar}J_{3}) \]
\[ H_{0} \equiv b \vec{x_{0}}.\vec{J} = b(\cos \theta J_{3}+\sin \theta
J_{1}) \]
\[ x_{0} \equiv (\theta ,\varphi = 0) \]
and $\vec{J}$ is the angular momentum operator. Note that due to
(7) , $H_{0}$ is time independent. Now , let $\mid \psi '> \equiv
U \mid \psi >$ and substitute \\ $\mid \psi >=U^{\dagger}\mid \psi '>$
in the Schr\"{o}dinger equation:
\[ i \hbar \partial _{t} \mid \psi >=H \mid \psi > \]
\begin{eqnarray}
i \hbar U \partial _{t}(U^{\dagger} \mid \psi '>) & = & UH(U^{\dagger}
\mid \psi '>) \nonumber \\
i \hbar \partial _{t}\mid \psi '> & = & (UHU^{\dagger}-i\hbar U\partial
_{t}U^{\dagger})\mid \psi '> \nonumber \\
& = & H'_{0} \mid \psi '>
\end{eqnarray}
where $H'_{0}$ is defined by :
\begin{eqnarray}
H'_{0} & \equiv & UHU^{\dagger}- i\hbar U\partial _{t} U ^{\dagger}
\nonumber \\
& = & H_{0}-\omega J_{3} \nonumber \\
& = & b[(\cos \theta -\omega /b)J_{3}+\sin \theta J_{1}] \nonumber
\end{eqnarray}
Again , because of (7) , $H'_{0}$ is time independent and hence (9)
can be immediately integrated to give:
\be
\mid \psi '(t)>=e^{-\frac{it}{\hbar}H'_{0}}\mid \psi '(0)>.
\ee
Using the definition of $\mid \psi '>$ one has:
\be
\mid \psi (t)>=U^{\dagger} \mid \psi '(t)>=e^{-\frac{it\omega}
{\hbar}J_{3}} e^{-\frac{it}{\hbar}H'_{0}}\mid \psi (0)>
\ee

Next , we show that the cyclic state vectors are eigenvectors of
$H'_{0}$. Let $\mid \tilde{\psi }(0)>$ be an exact cyclic state vector
, i.e. \[ \mid \tilde{\psi}(T)>=e^{i\alpha } \mid \tilde{\psi }(0)> \]
where $T\equiv 2\pi / \omega $ is the period of a cycle and $\alpha
\in \IR $. Using (11) one then has:
\be
e^{-\frac{2\pi i}{\hbar}J_{3}} e^{-\frac{iT}{\hbar}H'_{0}}
\mid \tilde{\psi}(0)>=e^{i\alpha}\mid \tilde{\psi}(0)>.
\ee
The last equation, (12), says that $\mid \tilde{\psi}(0)>$ is an
eigenvector of the operator $\exp (-\frac{2\pi i}{\hbar}J_{3})
\exp (-\frac{iT}{\hbar}H'_{0})$. However , the two operators
$\exp (-\frac{2\pi i}{\hbar}J_3)$ and $\exp (-\frac{iT}{\hbar}
H'_{0})$ commute, hence $\mid \tilde{\psi}(0)>$ must be a simultaneous
eigenvector of these operators and so of $H'_{0}$.
Furthermore, since $H_{0}$ and $H'_{0}$
do not commute, $\mid \tilde{\psi}(0)>$ is not an energy eigenvector.
This justifies our first assumption, 1., namely that the cyclic
state vectors are eigenvectors of an operator $H'_{0}$.

The second requirement, 2., is also fulfilled. Let us define the map
\[ F:M=S^{2} \sokk S^{2}=M \]
by \footnote{These formulae are local expressions, valid only in the
upper patch of $S^{2}$ (excluding the South Pole), $\theta \neq \pi $.
To get the expressions for the lower patch , one can simply take
$\theta $ to $\theta - \pi $ , in this case $\theta \neq 0 $.} :
\[ \forall x=(\theta ,\varphi ) \in S^{2} \hspace*{.5in}
F(x) \equiv \tilde{x}=(\tilde{\theta},\tilde{\varphi}) \]
\be
\cos \tilde{\theta} \equiv \frac{b}{\tilde{\omega}}(\cos \theta -
\frac{\omega}{b}) \hspace{0.2in} ,\hspace{0.2in}
\sin \tilde{\theta} \equiv \frac{b}{\tilde{\omega}}\sin \theta \hspace{0.2in}
, \hspace{0.2in} \tilde{\varphi} \equiv \varphi
\ee
\[ \tilde{\omega} \equiv b\sqrt{(\frac{\omega}{b})^{2}-2(\frac{\omega}{b})
\cos \theta +1} \]
Then, it is easy to see that $F$ is a smooth function of $x \in S^{2}$,
\footnote{Except for $\omega = b$ in which case $F$ fails to be well-defined
at the north pole. We discuss this case in section 3.} and
\[ H'_{0}(x_{0})=\frac{\tilde{\omega}}{b}H_{0}(\tilde{x_{0}})=
\frac{\tilde{\omega}}{b}H_{0}(F(x_{0})). \]
Let us now define the following hermitian operator :
\[ \tilde{H}_{0}(x_{0}) \equiv (H_{0}oF)(x_{0}) \]
Since $\tilde{H}_{0}$ is a scalar multiple of $H'_{0}$, the cyclic
state vectors are also eigenvectors of $\tilde{H}_{0}$.Thus condition
1. is satisfied for $\tilde{H}_{0}$ too. Next we realize that we have
chosen a fixed coordinate frame in which the system is at $\varphi =0$
at initial time: $t=0$. Allowing for an arbitrary choice of coordinates
we define:
\be
\tilde{H}(x) \equiv H(\tilde{x}) = H(F(x)) = (HoF)(x)
\ee
The cyclic state vectors $\mid \tilde{\psi}(x) >$ are then eigenvectors of
$\tilde{H}(x)$ , $\forall x \in M=S^{2} .$\footnote{Note that here the
point $x=(\theta ,\varphi )$ is the starting point ($t=0$), hence
$\tilde{H}(x)$ is still time indepentent and has all the properties
of $\tilde{H}_{0}$.} Hence, $\tilde{H}$ satisfies both 1. and 2. .

The adiabatic limit is $\omega \ll b$ , in which $F$ approaches to the
identity map. The (adiabatic) Berry connection is given by
\cite{berry,bohm}:
\[ A=i<n,x \mid d \mid n,x> \]
where $\mid n,x>$ is a single-valued basis eigenvector of $H(x)$ ,
and \[ \mid n,x(t=0)> = \mid \psi (0)>  \simeq  \mid \tilde{\psi}(0)>. \]
Choosing  $\mid n,x>=\mid k,x>$ , $k$ being an eigenvalue of $\vec{x}(t).
\vec{J}$, one obtains \cite{bohm}:
\be
A=-k(1-\cos \theta ) d\varphi .
\ee
The corresponding Berry's curvature 2-form is then given by :
\be
\Omega =dA=-k\sin \theta  d\theta \wedge d\varphi .
\ee

The non-adiabatic B-S bundle $\tilde{\lambda}$ is pullback of A-A
bundle $\eta (\equiv \eta _{{\cal N} = 1})$ induced by
\be
\tilde{f}=foF.
\ee
Thus,
\be
\tilde{\lambda}=\tilde{f}^{\star}(\eta )=(foF)^{\star}(\eta )=
F^{\star}(f^{\star}(\eta ))=F^{\star}(\lambda ) .
\ee
Similarly , since $f$ and $\tilde{f}$ pullback the connections , one has:
\be
\tilde{A}=\tilde{f}^{\star}({\cal A})=F^{\star}(f^{\star}({\cal A}))=
F^{\star}(A).
\ee
In (19) $\cal A $ and $\tilde{A}$ stand for the canonical (Stiefel)
connection (on $\eta $) , and the non-adiabatic (Berry) connection
(on $\tilde{\lambda}$ ) , respectively. To find the (local) expression
for $\tilde{A}$ , one can directly use (19). The result is
\be
\tilde{A}=-k(1-\cos \tilde{\theta}) d\varphi .
\ee
An alternative method to calculate $\tilde{A}$ is to define the
geometric phase to be the difference between the total phase and
the dynamical phase. This is done in \cite{bohm}. The result is
identical with (20) but the derivation is much longer.
The corresponding curvature 2-form (to $\tilde{A}$) is  given by :
\begin{eqnarray}
\tilde{\Omega} & = & d\tilde{A} = -k\sin \tilde{\theta} d\tilde{\theta}
\wedge d\varphi
\nonumber \\
& = & -k(\frac{b}{\tilde{\omega}})^{3}(1-\frac{\omega}{b}\cos \theta )
\sin \theta  d\theta \wedge d\varphi .
\end{eqnarray}
The non-adiabatic Berry (geometric) phase angle is obtained as the
holonomy element:
\be
\gamma = \oint _{C} \tilde{A} = \int _{S} \tilde{\Omega}
\ee
where $C=\partial S $ is defined in (7).

At this point we would like to emphasize that although conditions 1.
and 2. of page 3 seem to be quite restrictive, the system of (6) is
certainly not the only case where they are satisfied. Indeed, (6) is
a member of a class of quantum systems whose Hamiltonians belong to
a semisimple Lie algebra \cite{anan-sto,wang}. The method presented
in this paper applies to these systems. This is quite transparent in
the analysis of \cite{wang}.

\section{TOPOLOGY OF $\lambda$ AND $\tilde{\lambda}$}

First we introduce some notation. Let $\nu \equiv \omega /b $ ,
$ z \equiv \cos \theta $ , $ \tilde{z} \equiv \cos \tilde{\theta} $
, also define :
\be
F_{0}(z,\nu ) \equiv \frac{z- \nu }{\sqrt{\nu ^{2}-2z\nu +1}}=
\tilde{z}.
\ee
Note that the only difference between (13) and (23) is that in (23)
we chose a different set of coordinates , namely $(z,\varphi ) $ and
$(\tilde{z},\varphi )$ instead of $(\theta ,\varphi )$ and
$( \tilde{\theta },\varphi )$ , respectively. In this notation (20)
and (21) become:
\be
\tilde{A}=-k(1-\tilde{z}) d\varphi
\ee
\be
\tilde{\Omega}=k  d\tilde{z} \wedge d\varphi =\frac{-kb^{3}
\sin \theta (1-\nu \cos \theta)}{(\nu ^{2}-2\nu \cos \theta +1)^{\frac{3}{2}}}
d\theta \wedge d\varphi
\ee

The principal bundle $\tilde{\lambda}=\tilde{f}^{\star}(\eta )$
is induced by :
\be
\tilde{f}(x)=(foF)(x)=f(F(x))=f(\tilde{x}).
\ee
Thus using (4) one has
\be
\tilde{f}(x)=\mid \psi _{\tilde{x}}><\psi _{\tilde{x}} \mid
\ee
where $\mid \psi _{\tilde{x}}>$ is an eigenvector of $H(\tilde{x})=
\tilde{H}(x)$. As mentioned above the topologies of $\lambda $ and
$\tilde{\lambda} $ are determined by the homotopy classes of $[f]$
and $[\tilde{f}]$ in $[M=S^{2},\IC P(\infty )]$ , respectively.
Alternatively ,one can use the fact that
\[ [M,\IC P(\infty )] \cong H^{2}(M,\iz ) \]
and look at the first Chern numbers $c_{1}$ ,which also classify all
$U(1)$-bundles \cite{ali-bohm,kostant}.

We first claim that:
\newtheorem{smallnu}{Claim}
\begin{smallnu}
 The bundles $\lambda $ and $\tilde{\lambda}$ have the same topology
for $\omega <b$ ($\nu <1$).
\end{smallnu}
We prove this result using two different methods:
\begin{itemize}
\item by studying the homotopy classes ;
\item by directly calculating the first Chern numbers.
\end{itemize}
We begin by the first method. Let $\epsilon \in \IR ^{+} $
, be arbitrarily small. Define the map ${\cal F}:[\epsilon ,\nu ] \times
S^{2} \sokk \IC P(\infty) $ by :
\[ {\cal F}(\xi ,x) \equiv f(F(x,\xi ))\hspace*{.5in},\forall x \in
S^{2}, \forall \xi \in [\epsilon ,\nu ] \]
where $F(x,\xi ) \equiv (F_{0}(z,\xi ),\varphi ) $ and $F_{0}$ is
defined in (23). We calim that $\cal F$
is the desired homotopy map , i.e.\ it has the following properties :
\begin{enumerate}
\item ${\cal F}(\epsilon ,x) \equiv f(F(x,\epsilon ))=f(x)$ , since
$\epsilon \ll 1 $.
\item ${\cal F}(\nu ,x) \equiv f(F(x,\nu ))=f(\tilde{x})=\tilde{f}(x)$
, this is obvious by (23) and (26).
\item ${\cal F}$ is continuous. To see this , note that ${\cal F}=foF$.
The continuity of $f$ is assured by the continuity of $H=H(x)$ , so
$\cal F$ is continuous if and only if $F$ is.
As is seen from (23), $F$ has a discontinuity at $(x=\mbox{North Pole},
\xi =1)$ , i.e. $(z,\xi )=(1,1) $.
Hence , by requiring $\nu $ to be less than 1
, i.e.\ $\omega <b $ ,we have $\xi <1 $ and $F$ is continuous on its
domain.
\end{enumerate}

The alternative method is to explicitly calculate the first Chern
numbers. Fortunately the integrals can be easily evaluated and the
result is
\be
\tilde{c}_{1} = c_{1} = -2k.
\ee
In (28) , $ \tilde{c}_{1} =\frac{1}{2 \pi } \int _{S^{2}} \tilde{\Omega}$
and $c_{1}=\frac{1}{2 \pi} \int _{S^{2}} \Omega $ stand
for the first Chern numbers of $\tilde{\lambda}$ and $\lambda$
respectively \cite{nash}.

If we choose $\nu \geq 1 $ , the map $F$ and hence $\cal F$ become
discontinuous in the full domain of $\xi \in (0,\nu ] $. This marks
the possibility of a change in the topology of $\tilde{\lambda} $.
Examining (23), we can prove that for $\nu >1 $ ,
$\tilde{f}$ is homotopic to a constant map, hence
\begin{smallnu}
 The bundle $\tilde{\lambda}$ undergoes a change of topology at
$\nu = 1 (\omega =b) $ , and it is a trivial bundle for
$\nu > 1 (\omega >b) $.
\end{smallnu}
To see this let us transform the domain $[\nu , \infty )$ to a
closed one by considering the homotopy map ${\cal G}:[0,\frac{1}{\nu}]
\times S^{2} \sokk \IC\ P(\infty )$ ,defined by :
\[ {\cal G}(\zeta ,x) \equiv f(G(x,\zeta )) \] with
\[ G(x ,\zeta )=\left \{ \begin{array}{ll}
F(x,\xi =1/\zeta ) & \mbox{for $ 0<\zeta \leq 1/\nu $ } \\
{\cal S}:\mbox{South Pole} & \mbox{for $ \zeta =0 $ }
\end{array}
\right. \]
It is easy to see that , for $\xi \geq \nu >1  $(i.e. $\zeta \leq
1/\nu )$ , $G$ and $\cal G$ are continuous. Furthermore,
\[ {\cal G}(1/\nu ,x) =f(F(x,\nu ))=f(\tilde{x})=\tilde{f}(x) \]
and  \[ {\cal G}(0,x) \equiv f({\cal S})=\mbox{constant}. \]
Thus , $\tilde{f}\approx $(a constant map).

It should be emphasized that the triviality of $\tilde{\lambda}$
does not imply that the holonomy vanishes. As far as a non-flat
connection exists on $\tilde{\lambda}$ the curvature $\tilde{\Omega}$
does not vanish , and the holonomy group is non-trivial. The only
difference is that now one can smoothly deform the Hamiltonian in
such a way that the phase vanishes.

Again , we could directly calculate the first Chern numbers.
In general , the integrals may not be as readily calculable
as they are in our case. The results are listed as (29) , and
confirm our homotopy analysis.
\be
\tilde{c}_{1}= \left\{ \begin{array}{ll}
-2k & \mbox{for $\nu <1 $ } \\
0 & \mbox {for $\nu >1 $ }
\end{array}
\right.
\ee

There are several remarkable consequences of (29). First of all,
it shows a direct relation between a topological invariant of
a $U(1)$-bundle and the eigenvalues of angular momentum.
Secondly, for $\nu <1$ the fact that $\tilde{c}_{1}$
is an integer, translates to $k$ being a half-integer.
One can also try to use (25) to calculate $\frac{1}{2 \pi}
\int_{S^{2}}\tilde{\Omega}$ even when
$\nu =1 $, which corresponds to the
{\bf critical frequency}: $\omega = b$. However, one must realize that
the function
\mbox{$ F:S^{2} \sokk S^{2}$} fails
to be single-valued at the north pole ($N$). Actually, $F$ maps
$N$ to the equator. This is a consequence of the fact that
\mbox{$F(\theta ,\varphi )=(\tilde{\theta},\varphi )$}, but at
$N$ , $\varphi $ is not fixed. For $\nu <1$ or $\nu >1$, $N$ is
mapped to itself or to the south pole. In both of these cases , the
indeterminacy in $\varphi $ does not cause any problem. But for
$\nu =1$ , the situation is different and $F$ is not well-defined
over whole of $S^{2}$. Thus, one can not use $F$ and hence $\tilde{f}$
to pullback $\tilde{\lambda }$ from $\eta $, because the
fibre over $N$ would not be unique. However, one must note that this
is not a counterexample to our method. For $\nu =1$ there is no
(well-defined) $F$ that satisfy both conditions : 1. and 2. of
section 1, and our construction does not apply \footnote{ A rather trivial
solution to this problem is to simply exclude $N$ from the parameter space.
In that case,
\[ F : M'\equiv S^{2}-\{N\} \okk S^{2} \]
is well-defined and can be used to pullback a bundle $\tilde{\lambda }'$
from $\lambda $. The bundle $\tilde{\lambda}'=F\mid _{M'}^{\star }
(\lambda )$ has obviously trivial topology. The geometric phase
is again identified with the holonomy of $\tilde{\lambda}'$.}.

\newpage
\section{CONCLUSION}

The study of the topology of fibre bundles encountered in the
subject of Berry phase, suggests many interesting relations
between quantum mechanics and algebraic topology. Among these is the
relation between spin and the topological invariants such as Chern numbers.
For the example considered, there is a critical frequency at which
a topological change of the bundle structure occurs.

\section{ACKNOWLEDGMENTS}

A.M. would like to thank E.Demircan, O.T.Turgut, S.Gousheh for
their patiently listening to his arguments, and R.Lopez-Mobilia
for his plotting the graph of $F_{0}$ which is not presented here.
Also we would like to thank R.Murray for carefully reading
the first draft and bringing some algebraic mistakes to our attention.

\end{document}